\documentclass[letterpaper,aps,prl,twocolumn,superscriptaddress,showpacs]{revtex4}

\usepackage{graphicx}
\usepackage{amsmath}

\begin{document}

\title{Structure and Mass Absorption of Hypothetical Terrestrial Black Holes}

\author{A. P. VanDevender}
\affiliation{%
Halcyon Molecular, 505 Penobscot Dr, Redwood City, CA 94063
}%
\email{aaron@vandevender.com}
\author{J. Pace VanDevender}
\affiliation{Sandia National Laboratories MS-0125, Albuquerque, NM 87185-0125, USA}

\date{\today}

\begin{abstract}
The prospect of mini black holes, either primordial or in planned experiments at the Large Hadron Collider, interacting with the earth motivate us to examine how they may be detected and the scope of their impact on the earth. We propose that the more massive of these objects may gravitationally bind matter without significant absorption. Since the wave functions of gravitationally bound atoms orbiting a black hole are analogous to those of electrons around a nucleus, we call such an object the Gravitationally Equivalent of an Atom (GEA). Mini black holes are expected to lose mass through quantum evaporation, which has become well accepted on purely theoretical grounds. Since all attempts to directly observe x-rays from an evaporating black hole have failed, we examine the possibility of the inverse test: search for extant mini black holes by looking for emissions from matter bound in a GEA. If quantum evaporation does not occur, then miniature black holes left over from the early universe may be stable, contribute to dark matter, and in principle be detectable through emissions associated with the bound matter. We show that small black holes---with masses below ${\sim}10^{12}$~kg---can bind matter without readily absorbing it into the black hole but the emissions are too weak to be detected from earth. 
\end{abstract}


\maketitle


Quantum evaporation of mini black holes \cite{hawking:mnras-152-75, zeldovich:sov-astron-a-10-602} has been so thoroughly studied theoretically that it has become a keystone \cite{carlip:RPPhys-64-885} to the development of theories of quantum gravity---even though quantum evaporation has not been verified experimentally or observationally.  Consequently, the need to test the theory has become urgent.  So far, attempts to directly observe the x-ray signature of the final stages of evaporation have been unsuccessful \cite{xu:hepnp-23-35}.  We examine a complementary approach---looking for small black holes whose existence would invalidate the quantum evaporation theory as it is currently formulated.  The search requires knowledge of the structure, possible observables, and an estimate of the frequency of possible observations.  We suggest a likely equilibrium that can be used to compute the approximate signature of terrestrial black holes and estimate the frequency of black holes hitting the earth if they compose a significant fraction of dark matter.

Planned experiments to produce mini black holes (MBH) \cite{hawking:mnras-152-75, zeldovich:sov-astron-a-10-602} in the laboratory \cite{dimopoulos:prl-87-161602,argyres:pl-b4411-96,giddings:jmp-42-3082, stoecker:ijmpD-XX-XXX} and recent studies that conclude that quantum evaporation \cite{hawking:cmp-43-199} of MBHs is still an open question \cite{unruh:PRD-71-024028, helfer:rpp-66-943} motivate our examining the structure of terrestrial black holes and their absorption of surrounding matter. We find that atoms bind gravitationally to mini black holes to form the Gravitational Equivalent of an Atom (GEA). The long lifetime of a GEA impedes the growth of the mini black hole and increases the large and conservative estimate \cite{giddings:prd-78-035009} of the time a mini black hole requires to absorb the earth. Suggestions that MBHs may comprise a significant portion of dark matter \cite{carr:prd-50-4853, helfer:rpp-66-943} leads to the same questions of the fate of the earth if MBHs become trapped as terrestrial black holes. 

Although the mass of laboratory produced black holes is expected to be $<10^{-23}$~kg \cite{dimopoulos:prl-87-161602,argyres:pl-b4411-96,giddings:jmp-42-3082, stoecker:ijmpD-XX-XXX}, the calculated formation of primordial MBHs under various assumptions favors MBHs with masses $<10^{6}$~kg \cite{carr:prd-50-4853,green:prd-56-6166}.  If quantum evaporation is applicable to MBHs, then primordial black holes of initial mass $<10^{12}$~kg should have evaporated by now\cite{hawking:cmp-43-199}. That the energetic signature of their final evaporation has not been observed \cite{xu:hepnp-23-35} suggests they were not created in large numbers or that they do not evaporate.  We consider the possibility that MBHs do not evaporate and investigate their structure and the absorption of surrounding matter.

We report on a new configuration for MBHs in which normal matter is bound by gravity to a black hole just as electrons are bound by electrostatic forces to the nucleus of an atom. Hawking \cite{hawking:mnras-152-75} reported the atom-like behavior of a charged MBH, although he did not treat an uncharged MBH that can also form an atom-like structure through the gravitational force.

While we often think of the gravitational force as being insignificant on the microscale compared to the electric force, that is not the case for MBHs.  The de Broglie wavelength of an atom (i.e. an electrostatically shielded nucleus) with atomic mass $m$ gravitationally bound to a MBH of mass $M$ is the same as the wavelength of an electron in a hydrogen atom for

\begin{equation}
M = \frac{m_ee^2}{4\pi\epsilon_0Gm^2}.
\end{equation}

$M$ ranges from ${\sim}10^{6}$~kg for hydrogen to ${\sim}10^{3}$~kg for silicon to ${\sim}300$~kg for iron. Since the mass $M$ of primordial black holes are thought to have been $<10^{6}$~kg, MBHs can bind normal matter with sufficient energy to exist on earth.  A GEA is an ensemble of charge-neutralized nuclei gravitationally bound to an uncharged MBH and is the likely equilibrium configuration of a MBH in the terrestrial environment.  

Although even approximate relativistic solutions of the Klein-Gordon equation for a spinless particle in a Schwarzschild metric is very complex \cite{elizalde:prd-37-2127}, the solutions of greatest relevance to the probable mass distribution of laboratory and primordial MBHs ($<10^6$~kg) are for non-relativistic particles that are $>10^{10}$ Schwarzschild radii from the black hole.  Therefore, the Schr\"odinger equation with a Newtonian metric is relevant and more generally understandable.


We consider only black holes with zero charge and zero angular momentum.  To understand how such a black hole may exist on earth without consuming the surrounding mass, consider a black hole that interacts strongly with its nearest neighbors only through gravity. The black hole attracts the nuclei of surrounding matter with a gravitational potential $-GMm/r$ for a mass $m$ at distance $r$ from the black hole of mass M. We write the Schr\"{o}dinger equation with this potential
\begin{equation}
i\hbar\frac{\partial\psi}{{\partial}t} = 
-\frac{\hbar^2}{2m}\nabla^2\psi-\frac{GMm}{r}\psi,
\end{equation}
and find the ground state solution's radial part to be
\begin{equation}
\psi(r) = 
\frac{1}{\sqrt{\pi}}\left(\frac{GMm^2}{\hbar^2}\right)^{\frac{3}{2}}\exp\left(\frac{-GMm^2r}{\hbar^2}\right).
\label{radial-part}
\end{equation}
From Equation \ref{radial-part} we find the radial expectation value $\langle 
r\rangle$ for the ground state (principal quantum number $n=1$) to be
\begin{equation}
\langle r \rangle = \frac{3 \hbar^2}{2 G M m^2}.
\label{expectation-value}
\end{equation}
Similarly, we find the energy level $E_1$ of the GEA ground state to be
\begin{equation}
E_1 = -\left(\frac{GM}{\hbar}\right)^2\frac{m^3}{2}.
\end{equation}

Thus far we have treated the black hole at the center of the gravitational 
potential to be a point mass; however, black holes are usually thought to be 
large objects which absorb all nearby matter. To determine the limits within which our analysis is applicable, we must determine what sizes of black holes may be 
accurately approximated by a point mass.

The Schwarzschild radius for a black hole is given by $R_s$
\begin{equation}
R_s = 2 G M/c^2.
\label{schwarzshield}
\end{equation}
We require that this radius be much smaller than the orbital size 
of the bound particles:
\begin{equation}
R_s \ll \langle r \rangle.
\label{capture-relation}
\end{equation}
Substitution of Equations \ref{expectation-value} and \ref{schwarzshield} into Equation \ref{capture-relation} and taking $m$ to be the mass of a proton gives
\begin{equation}
M \ll \left(\frac{\sqrt{3}}{2}\right)\frac{\hbar c}{Gm} \approx 2.5{\times}10^{11} {\rm kg},
\label{mass-thresh}
\end{equation}
as the threshold for inhibiting black holes from absorbing all surrounding matter and is the defining equation for the Gravitational Equivalent of an Atom (GEA).

That the Schwarzschild radius is so much smaller than the radius of the bound particles differentiates a black hole at the center of a GEA from astrophysical black holes. Given the possibility of abundant small black holes which do not consume all surrounding matter, we examine the case of such an object interacting with the earth. The concern that a terrestrial GEA might absorb the earth is similar to the early 20th century expectation that electrons orbiting a nucleus should radiate their energy away and fall into the nucleus.  Since the electron energy levels are quantized and the expectation value of the radius of the ground state is much larger than the radius of the nucleus, the probability of an electron being captured by the nucleus is vanishingly small.  Similarly, particles of mass $m$ are unlikely to fall into the black hole at the center of a GEA; however, those few that do could, in principle, provide energy for observable emissions.


With the basic structure of a GEA in mind, let us consider three domains of GEA masses: unbound, neutral, and plasma. The lightest GEAs are in the unbound domain where the ground-state binding energy is less than the thermal energy of the surrounding matter. In this domain, ($|E_1| < {\sim}0.03$~eV or $M<5{\times}10^3$~kg) particles may be scattered by the MBH, but they will not bind into discrete shells.  The laboratory produced, TeV black holes would fall into this domain.

MBHs somewhat heavier than this have a sufficiently strong gravitational potential to bind atoms at room temperature. The Hamiltonian of a nearby electron would be dominated by the electrostatic potential of any gravitationally bound ions. Such an electron would bind more closely to the ion, forming a neutral atom, than it would to the black hole at the core of the GEA, thus these intermediate weight GEAs tend to be electrostatically neutral.

The heaviest GEAs have a sufficient gravitational force to bind the nuclei to the black hole more closely than the electrons are bound to the nuclei.

While the binding energy of a single ion follows the same dependence on $M$ as with neutral GEAs, a GEA with many bound ions and electrons will acquire a net positive charge inside the radial expectation value for the ions as the ions and electrons are no longer exactly colocated. This net positive charge repels the ions away from the center and pulls the electrons closer until the two radii are nearly equal. We call this quasi-charge-neutral situation a plasma GEA. We can find the point where these distributions are balanced (for a large number of ions, $N$) by modeling the total potential that the ions and electrons experience as a function of the effective charge number $Z_{\it eff}$ created by the slightly more tightly bound ions. Including both gravitational and electrostatic components to the potential, we find the radius of an ion $\langle r_i \rangle$ of mass $m_i$ and charge number $Z_i$ to be
\begin{equation}
\langle r_i \rangle = \frac{3\hbar^2}{2m_i}\left(GMm_i-\frac{e^2Z_iZ_{\it eff}}{4\pi\epsilon_0}\right)^{-1},
\end{equation}
while the radius of an electron $\langle r_e \rangle$ would be
\begin{equation}
\langle r_e \rangle = \frac{3\hbar^2}{2m_e}\left(GMm_e+\frac{e^2Z_{\it eff}}{4\pi\epsilon_0}\right)^{-1},
\end{equation}
where $m_e$ is the mass of an electron. By equating $\langle r_i \rangle$ and $\langle r_e \rangle$ and solving for $Z_{\it eff}$ we find the value of $Z_{\it eff}$ that yields the quasi-neutral distribution
\begin{equation}
Z_{\it eff} = \frac{GM4\pi\epsilon_0(m_i^2-m_e^2)}{e^2(m_e+Z_im_i)}.
\label{zeff}
\end{equation}
This $Z_{\it eff}$ only applies when the ground-state energy $E_1$ due to the gravitational potential is larger than the electrostatic binding energy of an electron to the ion. This condition implies that
\begin{equation}
M > M^\dag \equiv \frac{e^2}{G4\pi\epsilon_0m_i}\sqrt{\frac{m_e}{m_i}}.
\end{equation}
To apply this condition to $Z_{\it eff}$ we define the function $\sigma_{M^\dag}(M)$ which is small for $M<M^\dag$ and is close to 1 for $M>M^\dag$ to act as a weighting function for $Z_{\it eff}$,
\begin{equation}
\sigma_{M^\dag}(M) \equiv \frac{M}{M^\dag + M}.
\end{equation}
Substituting the composite $Z_{\it eff}\sigma_{M^\dag}(M)$ for $Z_{\it eff}$ we find approximate expressions for the ground-state radius $\langle r_i \rangle$ and energy $E_1$ for plasma GEAs.
\begin{equation}
\langle r_i \rangle = \frac{3\hbar^2}{2GMm_i}\left(m_i-\frac{Z_i(m_i^2-m_e^2)}{m_e+Z_im_i}\sigma_{M^\dag}(M)\right)^{-1}
\end{equation}
\begin{equation}
E_1 = -\frac{G^2M^2m_i}{2\hbar^2}\left(m_i-\frac{Z_i(m_i^2-m_e^2)}{m_e+Z_im_i}\sigma_{M^\dag}(M)\right)^2
\end{equation}

\begin{figure}
\includegraphics[width=3.25in]{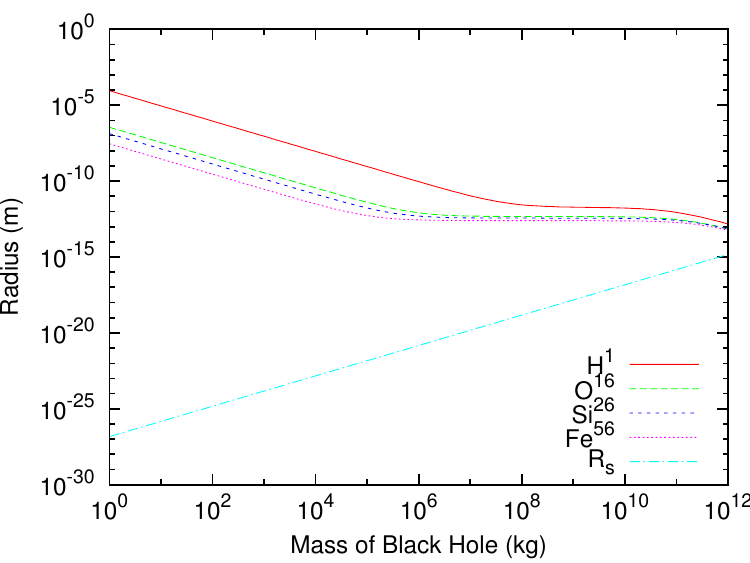}
\caption{Schwarzschild radius and the expectation value for the radial positions for H$^1$, N$^{14}$, Si$^{26}$, and Fe$^{56}$ as a function of the mass M of the central black hole.}
\label{ground-state-radius}
\end{figure}

For iron (the heaviest of the common planetary elements) with $m=56$ atomic mass units, the black hole mass at which the ground state radius equals the Schwarzschild radius is $M{\sim}10^{12}$~kg.  For comparison, astrophysical black holes created well after the formation of the universe have a threshold of about three solar masses or $M {\sim}6{\times}10^{30}$~kg. The radial expectation value $\langle{r}\rangle$ for all normal matter, including electrons, bound to such a black hole is well within the Schwarzschild radius, so matter is readily drawn into astrophysical black holes. The Schwarzschild radius for various MBH masses is shown in Figure \ref{ground-state-radius}. Since the expectation values of the ground-state radii are all vastly greater than a Plank length, quantum gravity is not required to examine the major features of this model.
\begin{figure}
\includegraphics[width=3.25in]{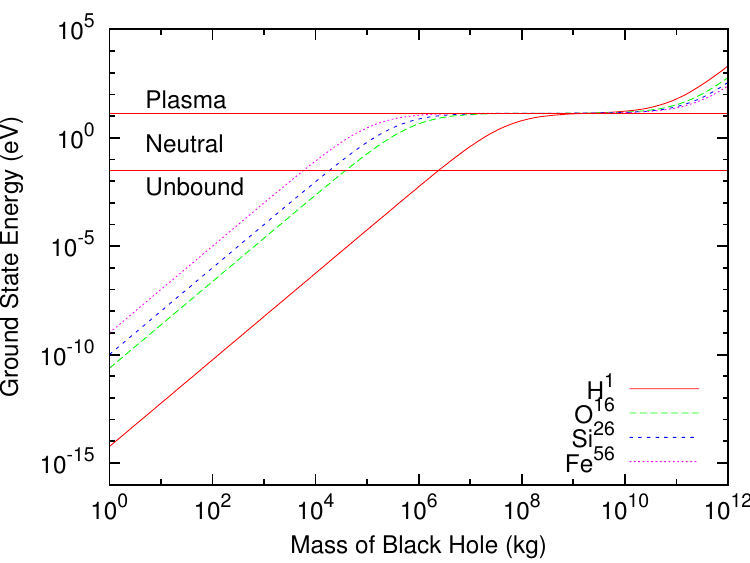}
\caption{Comparison of the ground state energy levels of H$^1$, N$^{14}$, Si$^{26}$, and Fe$^{56}$ as a function of 
mini black hole mass across the unbound, neutral, and plasma regions of GEAs.}
\label{ground-state-energy}
\end{figure}

\begin{figure}
\includegraphics[width=3.25in]{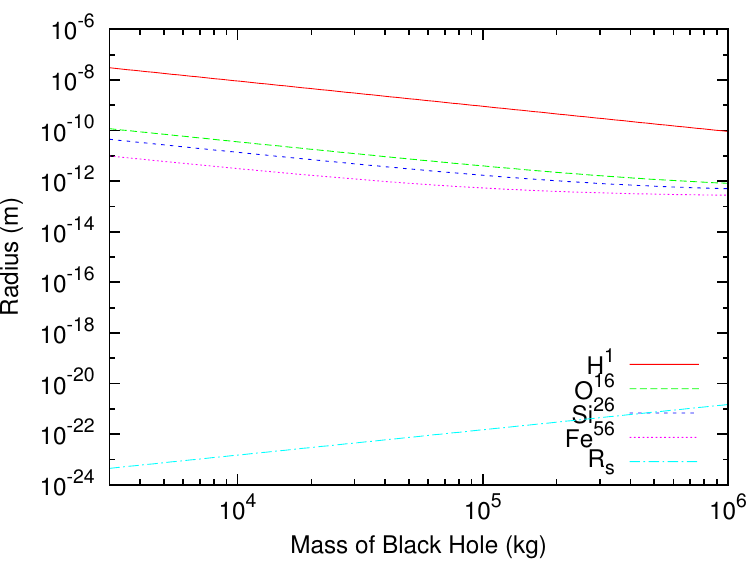}
\caption{Expansion of the ground state radius plot (Figure \ref{ground-state-radius})}
\label{ground-state-radius-big}
\end{figure}

\begin{figure}
\includegraphics[width=3.25in]{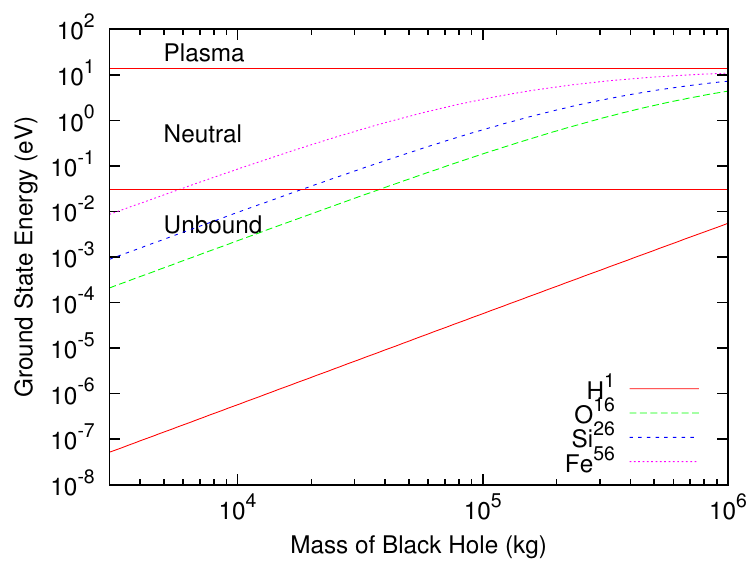}
\caption{Expansion of the ground state energy plot (Figure \ref{ground-state-energy})}
\label{ground-state-energy-big}
\end{figure}

The energy levels of a few common elements as a function of $M$ are shown in Figure \ref{ground-state-energy}. Terrestrial MBHs with $M >3,000$~kg should have bound neutrals that may be detectable and MBHs with $M >10^{5}$~kg could have detectable electromagnetic emissions from the bound plasma. Figures \ref{ground-state-radius-big} and \ref{ground-state-energy-big} provide a higher resolution version of Figures \ref{ground-state-radius} and \ref{ground-state-energy} for the region of greatest interest: 3,000~kg for the minimum mass with bound neutrals and $10^6$~kg for the maximum mass for primordial black holes \cite{carr:prd-50-4853,green:prd-56-6166}.


We note by setting $Z_{\it eff}$ equal to $\alpha^{-1}$ in equation \ref{zeff}, where $\alpha$ is the fine structure constant, we can determine the largest GEA for which the Dirac equation remains valid ($M_\alpha$):
\begin{equation}
M_\alpha=\frac{e^2(m_e+Z_im_i)}{\alpha G4\pi\epsilon_0(m_i^2-m_e^2)}.
\end{equation}

Substituting values for iron in $m_i=56$ and $Z_i=26$, we find $M_\alpha \approx 1.3\times 10^{11} {\rm kg}$, which is just beyond the plasma transition (Figure \ref{ground-state-energy}). Therefore, we do not expect any of the spontaneous pair creation processes associated with atoms of $Z>137$ to affect GEAs in the unbound or neutral regimes.


An analogous classical system of charge-neutralized ions bound by a central force is unstable \cite{krall-trivelpiece:482, alexeff:prl-45-354,chernin:pf-27-2319,stenzel:prl-60-704}.  Plasmas that do not have a Maxwellian velocity distribution in all three dimensions are prone to be unstable classically to anisotropy instabilities. If quantum effects do not preclude instability growth, such unstable dynamics could lead to a detectable radiation signature. Additional theoretical work is needed to understand the conversion process for GEAs and to identify the most energetic instabilities, the characteristic emission spectra, saturation mechanisms of the instabilities, and power level of the emissions that could be used to detect GEAs. 

As described above, their mass is likely to be between 10 and $10^6$~kg and their speed relative to the earth is likely comparable to the $2{\times}10^5$~m/s speed of the earth relative to the galactic center.  If the dark matter is primarily composed of MBHs of mass $M$ and is evenly distributed throughout our ${\sim}10^{42}$~kg galaxy, then the earth would be hit ${\sim}(4{\times}10^7 {\rm kg})/M$ times a year.  The MBHs would likely be detectable by strong electromagnetic emissions only if their surrounding matter is bound sufficiently strongly to support plasma instabilities, which implies a mass of $>10^5$~kg according to Figure \ref{ground-state-energy-big}. Under these conditions, about ${\sim}400$ MBHs per year would, in principle, be detectable through their strong electromagnetic emissions.  They may be detectable by more subtle emissions if the mass is $>3{\times}10^2$~kg, which corresponds to ${\sim}10^5$ per year.  Either number is sufficient to motivate a modest search of emissions from nearby space.

We live at the bottom of a gravitational well, so mini black holes reaching the earth will have a velocity $v$ at least the escape velocity from earth---11.2~km/s.  The scattering cross section with a bare black hole of a fermion (mass $m$) at low energy ($\hbar\gg{mvR_s}$) $\pi R_s^2/2$ \cite{doran:prd-71-124020}. Since the Schwarzschild radius for mini black holes of mass $< 10^6$~kg gives $R_s < 1.5{\times}10^{-21}$~m, the corresponding cross section for absorbing a fermion is $<3.5{\times}10^{-42}$~${\rm m}^2$ and the mean free path for scattering one electron in terrestrial matter is $>{\sim}200,000$ times the diameter of the earth. Furthermore, the black hole has a factor of $M/m_e$ greater momentum in the earth frame than the electron in the black-hole frame, so the mean free path for slowing down the black hole is $M/m_e$ greater or $2{\times}10^{39}$ to $2{\times}10^{41}$ earth diameters for black hole mass $M$ between $10^4$ and $10^6$~kg. Bare black holes do not interact with the earth.

Gravitationally bound particles in the GEA configuration mediate a stronger coupling as the bound particles interact with their environment and the momentum change is transferred to the mini black hole by the gravitational attraction. In the mass range of greatest interest between $10^4$ to $10^6$~kg, the radial expectation value for $N$ iron nuclei (atomic mass $A=56$ and atomic number $Z_i=26$) is $3{\times}10^{-12}$~m to $3{\times}10^{-13}$~m respectively. The size of the GEA is therefore much less than the ${\sim}2{\times}10^{-10}$~m inter-molecular spacing of water molecules or silicon atoms in the earth, so the GEA interacts with one molecule at a time. The electron-neutral scattering cross section is ${\sim}5{\times}10^{-15}$~${\rm m^2}$ or a factor of $10^{27}$ larger than that of a bare black hole.  Therefore, a GEA interacts with the surrounding neutral atoms and molecules much more strongly than does a bare black hole and might be detectable in the very unlikely event a GEA passes through a detector.

If a particle approaches a GEA it may be captured, it may scatter elastically, or it may strip an already bound particle off. The process of a GEA being captured is analogous to an electron being captured by an atom, though as the potential is mediated by gravitational rather than electromagnetic forces, conservation of energy is obtained by releasing a graviton rather than a photon. As the kinetic energy a MBH with initial velocity of ${\sim}11$~km/s is much larger than the total binding energy for even a very large number of atoms bound to the GEA, it will be quickly stripped of all mass as pass though the earth unimpeded as a bare MBH. Therefore, a search for electromagnetic signals from a GEA should focus on fast moving, unidentified rf sources in the space surrounding earth.  

Giddings and Mangano \cite{giddings:prd-78-035009} have assumed a worst case analysis of unimpeded accretion of mass by a mini black hole created in the earth's gravitational well and find that the time to assimilate the earth is much longer than the lifetime of the earth.  If a GEA with mass as large as $10^7$~kg were found in space and brought to the earth by human intervention and then escaped, the conservative assumptions of Giddings and Mangano yield an unimpeded assimilation time of $>10^{13}$ years for four dimensional space-time---more than 700 times the age of the universe.  The time is longer for smaller masses.  The quantum mechanics of the GEA impede the assimilation process and increase even that long assimilation time.


The time $\tau$ for a nucleus that is bound to a mini black hole of mass $M$ in the GEA to be absorbed by the central black hole is approximated as a particle flux  $|\psi(R_s)|^2c$ and an absorption cross section $4\pi R_s^2$ by\cite{flambaum:prd-63-084010, doran:prd-71-124020}
\begin{equation}
1/\tau \approx |\psi(R_s)|^2 4\pi R_s^2 c
\end{equation}
in which $|\psi(R_s)|^2$ is the probability density of the bound nucleus's wave function evaluated at the Schwarzschild radius and c is the speed of light in vacuum.

$\psi(r)$ is the solution to the Dirac equation for an electrostatically bound electron in a hydrogen atom \cite{bjorken-1998}, generalized to describe a combined gravitationally and electrostatically bound nucleus in a GEA by making the following substitutions:
\begin{eqnarray}
\alpha \mapsto& Z_i \alpha \\
Z \mapsto& Z_{\rm GEA}
\end{eqnarray}
in which $\alpha=1/137$, and $Z_{\rm GEA}=GMm4\pi\epsilon_0/(Z_ie^2)-Z_{\rm eff}$. The solution becomes
\begin{equation}
\begin{split}
|\psi(R_s)|^2=&\left(\frac{2}{\pi\Gamma(1+2\gamma)}\right)\left(\frac{cm_iZ_{\rm GEA}}{\hbar}\right)^3\left(Z_i\alpha\right)^{1+2\gamma} \times\\
&\left(\frac{2cm_iZ_{\rm GEA}R_s}{\hbar}\right)^{2(\gamma-1)} \times\\
&\exp(-2cm_iZ_{\rm GEA}Z_i{\alpha}R_s/\hbar),
\end{split}
\end{equation}
where
\begin{equation}
\gamma\equiv\sqrt{1-Z_{\rm GEA}^2Z_i^2\alpha^2}.
\end{equation}

We then assume that $M \ll \hbar c/(G m_i)$ (Eq. \ref{mass-thresh}). This implies that $R_sm_i \ll \hbar/c$, and by assuming $Z_{\rm GEA}Z_i\alpha \ll 1$ we approximate $\gamma \approx 1$. By applying these approximations, we obtain
\begin{equation}
\tau = \frac{c^3}{16{\pi}M^5B^3G^2},
\end{equation}
where
\begin{equation}
B\equiv\left(\frac{cG4\pi\epsilon_0m_iZ_i\alpha}{\hbar e^2}\right)\left(\frac{m_i}{Z_i}-\frac{m_i^2-m_e^2}{m_e+Z_im_i}\right).
\end{equation}
For example, $\tau \sim 10^{33}$ years for $M = 1$ kg and is much larger for the smaller black holes that might be formed in the LHC experiment.

Since the ground state radius is less than the ground state radius of normal atoms, only one nucleus \cite{giddings:prd-78-035009} can be absorbed at a time.  Assuming one bound nucleus is absorbed in time $\tau$, we compute the mass absorption rate for absorbing nuclei of atomic number $A$ as 
\begin{equation}
dM/dt {\sim}A m_p / \tau,
\end{equation}
where $m_p$ is the mass of a proton. Solving for the time $t_{GEA}$ required to grow from a mass $M_o$ to a mass $M_f \gg M_o$, we find that 
\begin{equation}
t_{\rm GEA}{\sim}\frac{c^3}{64Am_p{\pi}B^3G^2}(1/M_o^4 - 1/M_f^4)\\
\sim\frac{c^3}{64Am_p{\pi}B^3G^2M_o^4}
\end{equation}

The total time to consume a significant fraction of the earth is given by
\begin{equation}
t_{\rm total} =  t_{\rm GEA} + t_{\rm GM} + t_{\rm Bondi}.
\end{equation}

\begin{figure}
\includegraphics[width=3.25in]{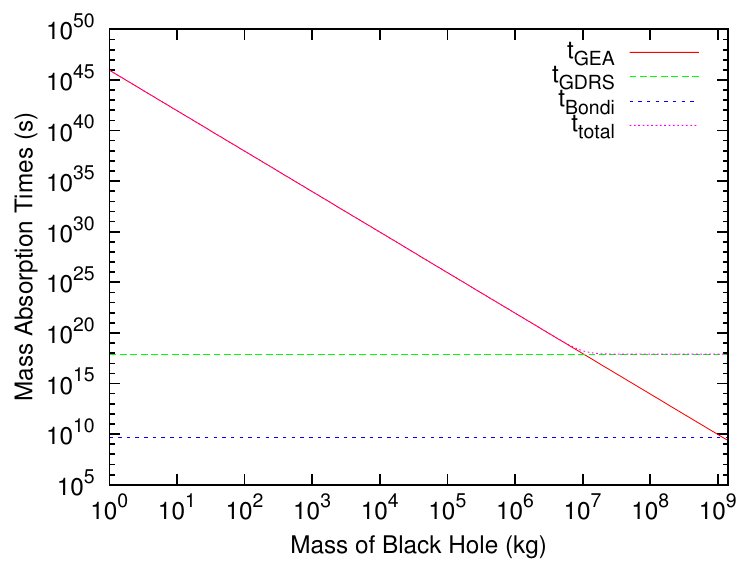}
\caption{Plot of mini black hole growth times.}
\label{growth-times}
\end{figure}

In this expression, $t_{\rm GM}$ is time to grow from $R_s \sim 10^{-12} m$ to $R_s \sim 10^{-10}$ m from Equation 4.20 of Giddings and Mangano\cite{giddings:prd-78-035009} and $t_{\rm Bondi}$  is the growth time by Bondi accretion from $R_s \sim 10^{-10}$~m to very large masses from Equation 4.41 of Giddings and Mangano, who conservatively  assume that the growth time ($t_{\rm GEA}$) of the mini black hole to nuclear dimensions---and hence to mass $M \sim 10^{14}$~kg---is negligible. The resulting growth times for 4-dimensional space time are compared as a function of $M_o$ in Figure \ref{growth-times}.  The GEA structure provides an even longer absorption time than the conservative estimate by Giddings and Mangano.

Giddings and Mangano also calculate the maximum assimilation time of a neutron star by a primordial black hole with initial mass of $10^{12}$~kg and find that the assimilation time is much less than the ${\sim}10^9$ year lifetime of neutron stars.  Therefore, the lifetime of neutron stars precludes the $10^{12}$~kg mini black holes as candidates for dark matter.  However, their same equations yield $3{\times}10^{16}$ and $3{\times}10^{14}$ years for mini black holes with initial mass of $10^4$ to $10^6$~kg respectively.  Therefore, the mini black holes of greatest interest are not precluded from being candidates for dark matter by the lifetime of a neutron star.

Finally, the power that could be available for electromagnetic emissions from a GEA is bounded by the gravitational energy per unit time of mass falling into the central black hole. The kinetic energy gain of a nucleus falling from the equilibrium radius $\langle{r}\rangle$ to $R_s =GMm (1/R_s - 1/\langle{r}\rangle) \sim mc^2$. This result and the mass absorption rate computed above gives the upper bound to the available power of $3{\times}10^{-24}\:{\rm (kg^5 W)}/M^{5}$. For the mass range of greatest interest for potentially observable primordial black holes ($10^4 - 10^6$~kg), the upper limit to the power available for electromagnetic emissions varies between ${\sim}10^{-29}$ and ${\sim}10^{-21}$~W which are not observable.


If black holes are created in the laboratory with the Large Hadron Collider and do not evaporate and/or if mini black holes were created in the formation of the universe and have survived the 13.7 billion years since the universe began, they should be present on or near earth.  In the former case, the binding energy for surrounding matter is too low to bind matter into quantum orbitals that might emit detectable radiation.  In the latter case, normal matter can be captured into quantum shells about the black hole to form the Gravitational Equivalent of an Atom (GEA).  In neither case will they absorb large amounts of mass very quickly.  

\begin{acknowledgments}
We gratefully acknowledge helpful conversations and constructive criticism from B. Carr, S. Carlip, J. Malenfant, V. V. Flambaum, M. Vahle, E. McGuire, K. Holley-Bockelmann, S. V. Greene, and D. Finley,
\end{acknowledgments}

\bibliography{references}
\bibliographystyle{apsrev}

\end{document}